\documentclass[lettersize,journal]{IEEEtran}
\usepackage[utf8]{inputenc}
\usepackage[T1]{fontenc}
\usepackage{amssymb}
\usepackage{amsmath}
\usepackage{amsthm}
\usepackage{mathrsfs}
\usepackage{graphicx}
\usepackage{array}
\usepackage{tabularx}
\usepackage{xcolor}
\usepackage{cite}
\usepackage{booktabs}
\usepackage{multirow}
\usepackage{url}
\usepackage{siunitx}
\usepackage{pdfpages}
\usepackage{balance} 
\usepackage{hyperref}
\usepackage[linesnumbered,ruled,vlined]{algorithm2e} 
\usepackage{algpseudocode}
\usepackage{enumerate}

\newtheorem{theorem}{Theorem}
\newtheorem{lemma}{Lemma}

\newtheorem{remark}{Remark}

\begin{document}

\title{Reinforcenment Learning-Aided NOMA Random Access: An AoI-Based Timeliness Perspective}

\author{Felippe~Moraes~Pereira, Jamil~de~Araujo~Farhat, João~Luiz~Rebelatto,~\IEEEmembership{Senior~Member,~IEEE,} Glauber~Brante,~\IEEEmembership{Senior~Member,~IEEE,} and Richard~Demo~Souza,~\IEEEmembership{Senior~Member,~IEEE.}%
\thanks{F. M. Pereira, J. A. Farhat, J. L. Rebelatto and G. Brante, CPGEI, UTFPR, Curitiba-PR, Brazil. felippepereira@alunos.utfpr.edu.br, \{jamilfarhat, jlrebelatto, gbrante\}@utfpr.edu.br.}%
\thanks{R. D. Souza, UFSC, Florianopolis/SC, Brazil. richard.demo@ufsc.br.}%
\thanks{This work was supported by CNPq and RNP/MCTIC (Brazil).}%
}

\maketitle

\begin{abstract}

In this paper, we investigate the age-of-information (AoI) of a power domain non-orthogonal multiple access (NOMA) network, where multiple internet-of-things (IoT) devices transmit to a common gateway in a grant-free random fashion. More specifically, we consider a framed setup composed of multiple time slots, and resort to the $Q$-learning algorithm to properly define, in a distributed manner, the time slot and the power level each IoT device transmits within a frame. In the proposed AoI-QL-NOMA scheme, the $Q$-learning reward is adapted with the aim of minimizing the average AoI of the network, while only requiring a single feedback bit per time slot, in a frame basis. Our results show that AoI-QL-NOMA significantly improves the AoI performance compared to some recently proposed schemes, without significantly reducing the network throughput. 

\end{abstract}


\section{Introduction} \label{sec:introduction}

Wireless sensor networks (WSNs) are crucial for the Internet of Things (IoT), providing real-time data for various applications. Accurate information from these sensors is essential for decision-making and actions in smart cities, healthcare, industrial automation, and environmental monitoring~\cite{Shafique.2020}. 

The basic building block of WSNs is based on a large number of sensors that communicate, interact and share data with each other. This challenging scenario is called massive IoT (mIoT), which focuses primarily on the efficient transmission of small amounts of data from many devices while accommodating other network constraints. Designing efficient, adaptable, and cost-effective massive IoT systems is a challenging task, given the increasing demands for IoT connectivity and the diverse application requirements~\cite{Jouhari.2023}.

In this context, the {\it age-of-information} (AoI) metric~\cite{Yates.IEEE_JSAC.2021} is crucial, as it ensures that data from WSNs remain timely and relevant. Unlike traditional metrics, AoI focuses on information freshness, vital for scenarios requiring immediate and accurate responses~\cite{Song.IoT.2024}. Keeping AoI low ensures actions are based on up-to-date information, enhancing the effectiveness of real-time monitoring, autonomous operations, etc.

The AoI metric in mIoT systems is linked to network access congestion. As random access (RA) grant-free (GF) techniques are limited by orthogonal resources, non-orthogonal multiple access (NOMA) strategies have gained prominence~\cite{Liu.2021}, allowing multiple users to communicate concurrently using the same time, frequency, or code resources.

Moreover, machine learning increasingly optimizes RA in IoT, where efficient resource allocation is vital for network performance and longevity~\cite{Yun.2021}. Among these options, $Q$-learning (QL) stands out as a powerful tool for resource allocation~\cite{Sutton.2018}, enabling networks to learn from their environment and make decisions that boost long-term rewards. This reinforcement learning approach allows networks to autonomously enhance their operations, improving adaptability to IoT needs.

Recent studies have examined AoI in RA~\cite{Bae.2022, Liu.2024, Ding.TWC.2023, Ding.WCL.2024}. Specifically,~\cite{Bae.2022} looks at AoI under a state-dependent slotted-ALOHA (SA) protocol, while~\cite{Liu.2024} suggests an AoI-oriented NOMA RA scheme with devices accessing the medium at different probabilities when AoI exceeds a threshold. From a grant-free perspective,~\cite{Ding.TWC.2023} investigates the impact of NOMA-assisted RA and OMA on AoI. Finally,~\cite{Ding.WCL.2024} evaluates the effect of different design strategies in NOMA-assisted RA.

Recently, QL-based methods have also been considered to improve GF access~\cite{Sharma.2019,Valente.2020,Liu.JSAC.2021,Jeong.2022,Cavalagli.2024}. In~\cite{Sharma.2019}, a QL-based method addresses network congestion with a reward based on the collision count per slot. Later,~\cite{Valente.2020} proposes a NOMA scenario requiring less feedback than~\cite{Sharma.2019}. Moreover,~\cite{Liu.JSAC.2021} examines QL methods for GF NOMA, focusing on traffic and techniques to reduce scheduling space and solve convergence issues.

While~\cite{Sharma.2019,Valente.2020,Liu.JSAC.2021} focus on throughput,~\cite{Jeong.2022,Cavalagli.2024} apply QL to improve AoI in WSNs. In~\cite{Jeong.2022}, the reward for successful transmission is proportional to the gain in freshness. In~\cite{Cavalagli.2024}, QL decides whether a device should transmit or remain silent, with the reward based on the device's AoI as in~\cite{Jeong.2022}. However, these works adopt a \textit{generate-at-will} policy where devices can generate new updates anytime, requiring time-slot feedback from the gateway (GW), which may be prohibitive for mIoT.

This work investigates a QL-assisted NOMA GF network in a framed scenario, where IoT device updates are generated under a \textit{generate-at-request} policy at the beginning of each frame~\cite{Ding.TWC.2023}. It aims to use QL in a distributed manner for IoT devices to choose their transmission time slots and power levels within a frame, reducing the network's AoI compared to other schemes. The contributions of this work are as follows:
\begin{itemize}
    \item We adopt $Q$-learning for time slot and power level allocation per IoT device within a frame, aiming at minimizing the average AoI of the network.
    \item We show that our proposed AoI-QL-NOMA scheme achieves near-optimal average AoI through a well-designed reward mechanism in the $Q$-learning algorithm, needing just one bit of feedback per time slot, sent at the end of the frame.
\end{itemize}

It is worthy emphasizing that the proposed AoI-QL-NOMA scheme differs from~\cite{Ding.TWC.2023}, which uses ALOHA-based repetitions, and also from~\cite{Cavalagli.2024}, which considers a \textit{generate-at-will} policy for time-slotted transmissions.

\section{Preliminaries} \label{sec:preliminaries}

\subsection{System Model} \label{ssec:system_model}

\begin{figure}[!t] 
\centering
\includegraphics[width=7cm]{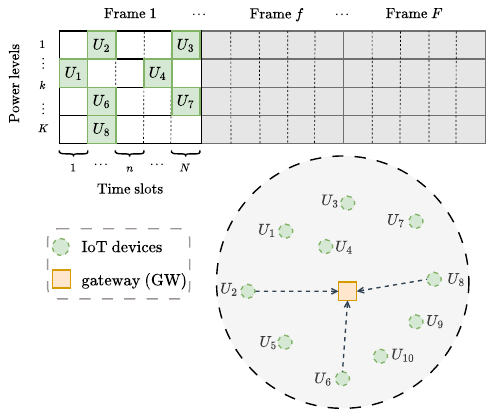}
\caption{System model: $M$ IoT devices communicate with a GW. Each frame contains $N$ time slots. NOMA-assisted RA is used, with $K$ power levels.}
\label{Fig:SystemModel}
\end{figure}

We consider the GF uplink of a WSN, where $M$ IoT devices labeled $\{U_m\}_{m=1}^{M}$ independently transmit to a common GW, as shown in Fig.~\ref{Fig:SystemModel}. Communication is time synchronous, being each of the $F$ frames composed of $N$ time slots.  

In addition, NOMA-assisted RA is employed, where the GW defines $K$ power levels $\{P_{k}\}_{k=1}^{K}$ to be reached at destination, which are established to ensure the success of the successive interference cancellation (SIC) technique at the receiver. Such levels are designed such that $\log\left(1+\frac{P_k}{1+\sum_{k'=k+1}^{K} P_{k'}}\right) = R$ for $1 \leq k \leq K-1$ and $\log(1+P_K)=R$, where $R$ is the transmission rate and the noise power is normalized to one~\cite{Ding.TWC.2023}. Therefore, NOMA enables up to $K$ devices to share the same time slot, as long as they adopt different power levels. However, this strategy requires the devices to have transmit channel state information (CSI)\footnote{One approach to obtain CSI is through the broadcast of pilot signals from the GW before data transmission from the IoT devices~\cite{Ding.TWC.2023}.} to properly adapt their transmit power and mitigate the fading, assumed to be Rayleigh distributed (independent and identically distributed between frames and devices).

Moreover, packet generation follows a {\it generate-at-request} model~\cite{Ding.TWC.2023}, where devices generate a new packet at the beginning of each frame, which is transmitted in a grid position $(k,n)$, where $\{k\}_{k=1}^{K}$ corresponds to the power level index, while $\{n\}_{n=1}^N$ represents the time slot index.  

\subsection{Age-of-Information (AoI)} \label{ssec:aoi}

The AoI metric quantifies the freshness of the information available at the GW. The instantaneous AoI of $U_m$ is defined as $\Delta_m(t) = t - r_m(t)$, which depends on the generation time $r_m(t)$ of the most recent packet from $U_m$ that was successfully delivered to the GW~\cite{Yates.IEEE_JSAC.2021}.

Being a stochastic process, the average AoI is obtained as
\begin{equation}
    \bar \Delta_m = \lim_{T \to \infty} \frac{1}{T} \int_{0}^{T} \Delta_m(t)\, \mathrm{d}t.
    \label{eq:avg_delta}
\end{equation}

\begin{figure}[!t] 
\centering
\includegraphics[width=5cm]{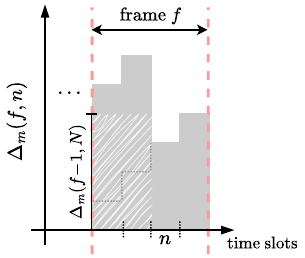}
\caption{AoI evolution within a frame. In this example, $U_m$ is decoded in time slot $n$. The sketched area represents the cumulative AoI that varies with $n$.}
\label{Fig:aoi_evolution_frame}
\end{figure}

In our setup, since the packets are generated at the beginning of each frame, $\Delta_m(f,n) = n$ if $U_m$ is recovered in time slot $n$ of frame $f$, and $\Delta_m(f,n)= \Delta_m(f,n\!-\!1) + 1$ otherwise. Also, in case $n=1$ it returns to the previous frame $\Delta_m(f\!-\!1,N) + 1$. In particular, note in Fig.~\ref{Fig:aoi_evolution_frame} that the sooner a device is recovered by the GW within a frame, the fresher its average AoI.  Remarkably, this differs from the throughput or reliability metrics, whose performance does not depend on $n$.

\subsection{$Q$-Learning} \label{ssec:q_learning}

The $Q$-Learning algorithm is a reinforcement learning technique particularly suited to address resource allocation problems in communication systems due to its effectiveness in solving dynamic decision-making problems~\cite{Sutton.2018}. This algorithm adjusts the selection strategy based on rewards. 

Given an agent indexed by $i$ that belongs to a state $S_{i}$ and performs an action $A_{i}$, with the goal of maximizing its reward, the updates in the $Q$-table are defined as follows~\cite{Sutton.2018}
\begin{equation}
Q(S_{i}, A_{i}) \leftarrow (1-\alpha)Q(S_{i}, A_{i}) + \alpha(\epsilon +\gamma \max_{a} Q(S_{i+1}, a)),
\label{eq:QTable}
\end{equation}
where $\alpha$ is the learning rate, $\gamma$ is the discount factor quantifying the importance of future rewards, and $\epsilon$ is the reward.

\section{Proposed AoI-QL-NOMA Scheme} \label{sec:ql_noma}

In this letter, we adopt $Q$-learning to properly allocate the time slot and power level for each IoT device in a decentralized manner, with the aim of reducing the AoI of the system.

In the proposed AoI-QL-NOMA scheme, the IoT devices act as agents of the $Q$-learning algorithm, managing their own $K \times N$ $Q$-table. The state-action pair $\phi_m(f)=(k,n)$ is composed of the time slot $n$ and the power level $k$ used by $U_m$ to communicate with the GW in frame $f$~\cite{Valente.2020}. 

Following~\cite{Valente.2020}, we consider that at the end of each frame, the GW feeds back to the IoT devices one bit per time slot (in a broadcast fashion, regardless the number of devices). Ommiting the frame index, let us refer to $b(n)\in\{{\tt0},{\tt 1}\}$ as the feedback bit associated with time slot $n$. In the QL-NOMA scheme from~\cite{Valente.2020}, such feedback bit simply indicates whether the packets transmitted within time slot $n$ have been correctly decoded by the GW via SIC or not. However, in the proposed AoI-QL-NOMA scheme, the GW defines such bit as\footnote{Note that~\eqref{eq:feedback_bit} assumes that the GW is capable of estimating $M$, which is a reasonable assumption since each IoT device transmits only a single time per frame. Then, the GW can estimate $M$ even in case of collisions.}   
\begin{equation} \label{eq:feedback_bit}
	b(n)=
    \begin{cases}
      {\tt 1}, & \text{if SIC successful {\bf and} } n \leq \lceil M/K \rceil \\
      {\tt 0} , & \text{otherwise,}
    \end{cases}
\end{equation}
where $\lceil \cdot \rceil$ is the ceil operation. Additionally, the dependence of $n$ differs from~\cite{Valente.2020}, where the feedback bit is fixed irrespective of the time slot . Thus, based on~\eqref{eq:feedback_bit}, each IoT device updates its own $Q$-table following~\eqref{eq:QTable} by considering the reward: 
\begin{equation} \label{eq:reward}
    \epsilon(k,n)= 2\,b(n)-1,
\end{equation}
{\it i.e.}, bit ${\tt 1}$ is mapped into a positive reward $(+1)$, while bit ${\tt 0}$ is converted in a negative reward ($-1$). The reason behind such a policy is to aggregate the IoT devices in the early time slots of the frame. Since up to $K$ devices are able to transmit concurrently, one needs $\lceil M/K \rceil$ time slots to accommodate all the $M$ IoT devices. Although this approach may not be relevant from a throughput perspective (it actually slightly reduces throughput), we will show that it considerably reduces AoI. In what follows, we present some results about the performance of the proposed AoI-QL-NOMA scheme, which are valid for a sufficiently large number of frames $F$.

\begin{lemma}
    An IoT device recurrently served in time slot $n$ has an average AoI given by
    \begin{equation} \label{eq:average_aoi_n}
        \bar{\Delta}_m(n) = \frac{N-1}{2} + n. 
    \end{equation}
\end{lemma}
\begin{IEEEproof}
    The proof follows by calculating the area under the instantaneous AoI from Fig.~\ref{Fig:aoi_evolution_frame}, assuming that $\Delta_m(f-1,N) = N$ (\textit{i.e.}, $U_m$ has been served in frame $f-1$). While the sketched part has area $N(n-1)$, the remaining staircase  has area $N(N+1)/2$. By adding the terms and normalizing by $N$, one has~\eqref{eq:average_aoi_n}, concluding the proof.  
\end{IEEEproof}

\begin{theorem}
    The average AoI of the proposed AoI-QL-NOMA scheme is lower bounded by
    \begin{equation} \label{eq: bound_aoi}
        \bar{\Delta}_{\text{QL-NOMA}} \geq \frac{N+\lceil M/K \rceil}{2},
    \end{equation}
    where $\lceil M/K \rceil \leq N$.
\end{theorem}
\begin{IEEEproof}
Note that $\lceil M/K \rceil$ is the minimum number of time slots necessary to accommodate all the $M$ devices. Thus, a natural bound would be obtained with the aid of~\eqref{eq:average_aoi_n} as
        \begin{equation*}
        \bar{\Delta}_{\text{QL-NOMA}} \geq \frac{1}{\lceil M/K \rceil} \sum_{n=1}^{\lceil M/K \rceil} \bar{\Delta}_m(n),
    \end{equation*}
    which, after expanded, reduces to~\eqref{eq: bound_aoi}, concluding the proof. 
\end{IEEEproof}

\begin{remark} 
    When $\lceil M/K \rceil = N$, the proposed AoI-QL-NOMA reduces to QL-NOMA with fixed reward from~\cite{Valente.2020}, whose average AoI $\bar{\Delta}_{\text{QL-NOMA}} \geq N$, as from~\eqref{eq: bound_aoi}. 
\end{remark}

\section{Numerical Results} \label{sec:numerical_results}

This section evaluates the performance of the proposed AoI-QL-NOMA scheme. We adopt as benchmark the following schemes: `SA-NOMA REPET.'~\cite{Ding.TWC.2023}, where devices transmit in a RA fashion per time slot with transmission probability $K/M$, until receiving an ACK from the GW. That is, retransmissions are allowed. The `SA-NOMA' is a particular case where each device randomly selects a single pair of time slot and power level to transmit within the frame. Finally, `QL-NOMA' refers to the scheme from~\cite{Valente.2020}, where the reward is fixed to $\epsilon=\{-1, 1\}$ in case of failure or success, respectively.

\begin{figure}[!t]
    \centering
    \includegraphics[width=0.45\textwidth]{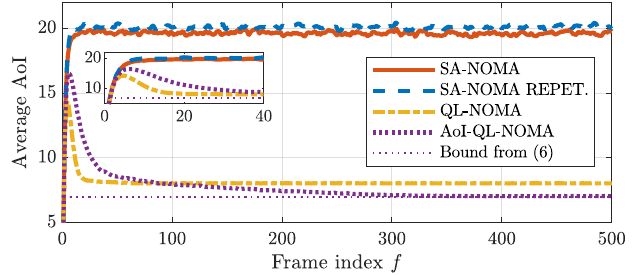}
    \caption{Average AoI versus frame index, for $N=8$, $K=3$, $M=18$.}
    \label{fig:AoI_per_frame}
\end{figure}

\begin{figure}[!t]
    \centering
    \includegraphics[width=0.48\textwidth]{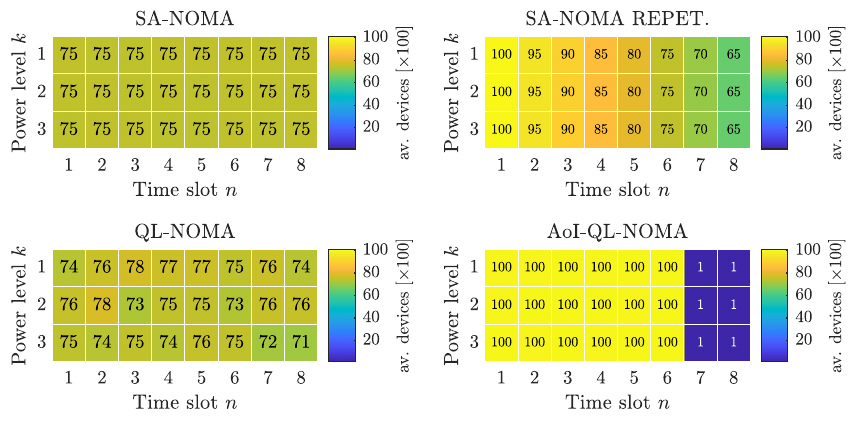}
    \caption{Average devices per grid ($\times 100$), for $N=8$, $K=3$, $M=18$.}
    \label{fig:grid_usage}
\end{figure}

\begin{figure}[!t]
    \centering
    \includegraphics[width=0.48\textwidth]{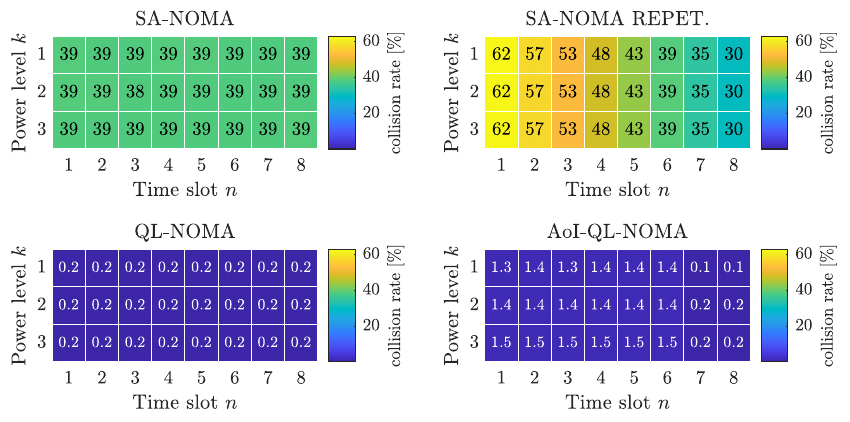}
    \caption{Average collision rate, for $N=8$, $K=3$, $M=18$.}
    \label{fig:grid_collision}
\end{figure}

Figs.~\ref{fig:AoI_per_frame}-\ref{fig:grid_collision} consider a scenario with $N=8$ time slots per frame, $K=3$ power levels, $N=18$ IoT devices, $Q$-learning parameters $\alpha=0.1$ and $\gamma=0.5$~\cite{Valente.2020}. Such results are the average of $10^3$ independent runs, each with $F=10^3$ frames.

Fig.~\ref{fig:AoI_per_frame} presents the evolution of the average AoI per frame versus the frame index $f$. One can see that, despite requiring more frames to converge, the proposed AoI-QL-NOMA scheme achieves a lower average AoI than all the other schemes, approaching the bound from~\eqref{eq: bound_aoi}  ($\approx 7$ time slots versus $8$ time slots achieved by the QL-NOMA with fixed reward, $12.5\%$ better), without requiring any additional resource neither in terms of feedback or processing capabilities. This improvement is achieved just by properly adjusting the QL reward as from~\eqref{eq:reward}, which allocates the IoT devices in the first time slots of the frame. The SA based schemes present similar performances, and interestingly the repetition-based scheme is the one that achieves the higher AoI value.   

The average grid usage is depicted in Fig.~\ref{fig:grid_usage}, which represents the average number of devices ($\times 100$) that selects each power-level/time-slot pair $(k,n)$. As expected, while SA-NOMA and QL-NOMA uniformly distribute the users throughout the grid, the devices tend to choose the initial time slots when operating under the proposed AoI-QL-NOMA scheme, being the two last time slots only chosen before convergence. The `SA-NOMA REPET.' scheme also concentrates more transmissions in the initial time slots, but, besides presenting on average at most one device per grid slot for small values of $n$, its random nature (without any learning process) leads to an increased number of collision, as depicted in Fig.~\ref{fig:grid_collision}.

Fig.~\ref{fig:grid_collision} presents the average collision rate (in \%) for each $(k,n)$ pair of the grid. The SA-NOMA scheme with repetitions shows a considerably high collision rate, exceeding 50\% initially, compromising its performance. In contrast, QL-NOMA has an extremely low and uniform collision rate, due to faster convergence. The proposed AoI-QL-NOMA scheme has a slightly higher collision rate than the fixed reward counterpart for $n \leq 6$, due to device positioning, temporarily increasing collisions before convergence. Hence, the proposed AoI-QL-NOMA scheme may have slightly lower throughput compared to QL-NOMA with fixed reward. For instance, the normalized throughput for the proposed and QL-NOMA schemes are $0.974$ and $0.991$, respectively, a decrease of only $1.6$\%. This throughput penalty vanishes as the number of frames increases.

\begin{figure}[!t]
    \centering
    \includegraphics[width=0.48\textwidth]{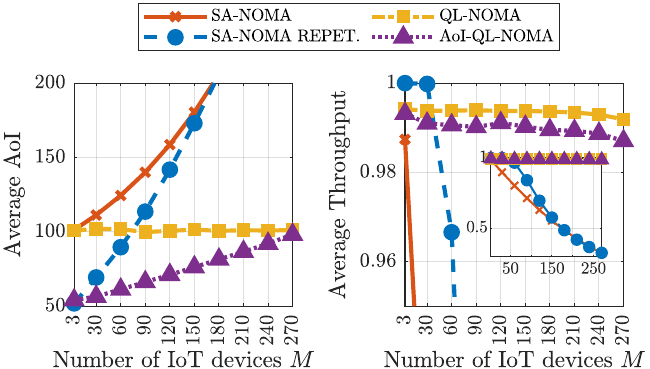}
    \caption{Average AoI (left) and Average Throughput (right) as a function of the number of IoT devices $M$.}
    \label{fig:AoI_throughput_users}
\end{figure}

The small number of devices in previous figures balances computational complexity and highlights the AoI-QL-NOMA protocol's mechanisms. To complement the analysis, Fig.~\ref{fig:AoI_throughput_users} shows the average AoI of the network (left) against the number of devices $M$, considering $K=3$, $N=100$, and learning parameters $\alpha=0.1$, $\gamma=0.5$. Results consider $F=10^4$ frames over $10$ runs. The proposed scheme outperforms others across the range, with gain decreasing as users increase. When fully saturated {\it i.e., $M=KN$}, AoI-QL-NOMA reduces to fixed reward QL-NOMA. Fig.~\ref{fig:AoI_throughput_users} also shows the average throughput (right), indicating minimal throughput loss with the proposed AoI-QL-NOMA, regardless of device numbers.

\section{Conclusion} \label{sec:final_comments}

We propose a distributed $Q$-learning approach to allocate pairs of time slots and power levels in a NOMA-based IoT network, with the aim of reducing the average AoI of the network. Numerical results show that the proposed QL-NOMA scheme outperforms the benchmark SA-NOMA approach under various circumstances. Future works include, for example, the proposal of an age-dependent policy, where only devices above a given age threshold are allowed to access the channel. 

\bibliographystyle{IEEEtran}
\bibliography{IEEEabrv,references}

\begin{thebibliography}{10}
\providecommand{\url}[1]{#1}
\csname url@samestyle\endcsname
\providecommand{\newblock}{\relax}
\providecommand{\bibinfo}[2]{#2}
\providecommand{\BIBentrySTDinterwordspacing}{\spaceskip=0pt\relax}
\providecommand{\BIBentryALTinterwordstretchfactor}{4}
\providecommand{\BIBentryALTinterwordspacing}{\spaceskip=\fontdimen2\font plus
\BIBentryALTinterwordstretchfactor\fontdimen3\font minus
  \fontdimen4\font\relax}
\providecommand{\BIBforeignlanguage}[2]{{%
\expandafter\ifx\csname l@#1\endcsname\relax
\typeout{** WARNING: IEEEtran.bst: No hyphenation pattern has been}%
\typeout{** loaded for the language `#1'. Using the pattern for}%
\typeout{** the default language instead.}%
\else
\language=\csname l@#1\endcsname
\fi
#2}}
\providecommand{\BIBdecl}{\relax}
\BIBdecl

\bibitem{Shafique.2020}
K.~Shafique, B.~A. Khawaja, F.~Sabir, S.~Qazi, and M.~Mustaqim, ``Internet of
  things ({IoT}) for next-generation smart systems: A review of current
  challenges, future trends and prospects for emerging {5G-IoT} scenarios,''
  \emph{IEEE Access}, vol.~8, pp. 23\,022--23\,040, 2020.

\bibitem{Jouhari.2023}
M.~Jouhari, N.~Saeed, M.-S. Alouini, and E.~M. Amhoud, ``A survey on scalable
  {LoRaWAN} for massive {IoT}: Recent advances, potentials, and challenges,''
  \emph{{IEEE} Commun. Surveys Tuts.}, vol.~25, no.~3, pp. 1841--1876, 2023.

\bibitem{Yates.IEEE_JSAC.2021}
R.~D. Yates, Y.~Sun, D.~R. Brown, S.~K. Kaul, E.~Modiano, and S.~Ulukus,
  ``{A}ge of {I}nformation: An introduction and survey,'' \emph{{IEEE} J. Sel.
  Areas Commun.}, vol.~39, no.~5, pp. 1183--1210, 2021.

\bibitem{Song.IoT.2024}
T.~Song and Y.~Kyung, ``Deep reinforcement learning based
  {Age-of-Information}-aware low-power active queue management for {IoT} sensor
  networks,'' \emph{{IEEE} Internet Things J.}, pp. 1--1, 2024.

\bibitem{Liu.2021}
J.~Liu, G.~Wu, X.~Zhang, S.~Fang, and S.~Li, ``Modeling, analysis, and
  optimization of {G}rant-{F}ree {NOMA} in massive {MTC} via stochastic
  geometry,'' \emph{{IEEE} Internet Things J.}, vol.~8, no.~6, pp. 4389--4402,
  2021.

\bibitem{Yun.2021}
W.-K. Yun and S.-J. Yoo, ``{$Q$}-{L}earning-based data-aggregation-aware
  energy-efficient routing protocol for wireless sensor networks,'' \emph{IEEE
  Access}, vol.~9, pp. 10\,737--10\,750, 2021.

\bibitem{Sutton.2018}
R.~S. Sutton and A.~G. Barto, \emph{Reinforcement Learning: An Introduction},
  2nd~ed.\hskip 1em plus 0.5em minus 0.4em\relax The MIT Press, 2018.

\bibitem{Bae.2022}
Y.~H. Bae and J.~W. Baek, ``{A}ge of {I}nformation and throughput in random
  access-based {IoT} systems with periodic updating,'' \emph{{IEEE} Wireless
  Commun. Lett.}, vol.~11, no.~4, pp. 821--825, 2022.

\bibitem{Liu.2024}
Y.~Liu, L.~X. Cai, Q.~Chen, H.~Zhang, F.~Hou, and T.~H. Luan, ``Minimizing age
  of information in nonorthogonal random access networks,'' \emph{{IEEE}
  Internet Things J.}, vol.~11, no.~14, pp. 24\,886--24\,902, 2024.

\bibitem{Ding.TWC.2023}
Z.~Ding, R.~Schober, and H.~V. Poor, ``Impact of {NOMA} on {A}ge of
  {I}nformation: A {G}rant-{F}ree transmission perspective,'' \emph{{IEEE}
  Trans. Wireless Commun.}, vol.~23, no.~5, pp. 3975--3989, 2024.

\bibitem{Ding.WCL.2024}
------, ``{NOMA}-assisted grant-free transmission: How to design pre-configured
  {SNR} levels?'' \emph{{IEEE} Wireless Commun. Lett.}, vol.~13, no.~2, pp.
  412--416, 2024.

\bibitem{Sharma.2019}
S.~K. Sharma and X.~Wang, ``Collaborative distributed {Q}-learning for {RACH}
  congestion minimization in cellular {IoT} networks,'' \emph{{IEEE} Commun.
  Lett.}, vol.~23, no.~4, pp. 600--603, 2019.

\bibitem{Valente.2020}
M.~V. da~Silva, R.~D. Souza, H.~Alves, and T.~Abrão, ``A {NOMA}-based
  {$Q$}-learning random access method for machine type communications,''
  \emph{{IEEE} Wireless Commun. Lett.}, vol.~9, no.~10, pp. 1720--1724, 2020.

\bibitem{Liu.JSAC.2021}
J.~Liu, Z.~Shi, S.~Zhang, and N.~Kato, ``Distributed {$Q$}-{L}earning aided
  uplink grant-free {NOMA} for massive machine-type communications,''
  \emph{{IEEE} J. Sel. Areas Commun.}, vol.~39, no.~7, pp. 2029--2041, 2021.

\bibitem{Jeong.2022}
M.~Jeong, G.~Seo, and E.~Hwang, ``Age of information optimization by deep
  reinforcement learning for random access in machine type communication,'' in
  \emph{IEEE Int. Conf. on Big Data}, 2022, pp. 6670--6672.

\bibitem{Cavalagli.2024}
C.~Cavalagli, L.~Badia, and A.~Munari, ``Reinforcement learning for age of
  information aware transmission policies in slotted {ALOHA} channels,'' in
  \emph{19th Int. Symp. Wireless Commun. Syst. (ISWCS)}, 2024, pp. 1--6.

\end{thebibliography}

\end{document}